\documentclass[11pt,a4paper]{article}
\pdfoutput=1    
\usepackage{jheppub}

\usepackage{amsmath}
\usepackage{amssymb}
\usepackage{mathrsfs}
\usepackage{enumerate}

\renewcommand{\thefigure}{\arabic{figure}}

\newtheorem{theorem}{Theorem}[section]
\newtheorem{lemma}[theorem]{Lemma}
\newtheorem{definition}[theorem]{Definition}
\def\proof{\noindent{\em Proof.~}}

\def\eq#1 { \begin{equation} #1 \end{equation} }
\def\eqn#1{ \begin{eqnarray} #1 \end{eqnarray} }
\def\nn { \nonumber }
\def\a{{\alpha}}

\def\l{{\lambda}}

\def\Lie{\mathcal{L}}
\def\cM{{\mathcal{M}}}
\def\cO{{\mathcal{O}}}
\def\vx{{\vec{x}}}
\def\vy{{\vec{y}}}
\def\ph#1{\phantom{ #1 }}
\def\d{{\partial}}

\def\Im{{\rm Im}\,}
\def\Reals{{\mathbb{R}}}

\def\Com#1#2{\left[ #1 , #2 \right]}
\def\C#1{\left\langle #1 \right\rangle}

\title{On higher spin symmetries in de Sitter QFTs}

\author[a,b]{Renato Costa}

\author[b]{Ian A.~Morrison}

\affiliation[a]{Instituto de F\'isica Teorica, UNESP-Universidade
  Estadual Paulista, \\
  R.~Dr.~Bento T.~Ferraz 271, Bl.~II, Sao Paulo 01140-070, SP, Brazil}

\affiliation[b]{Department of Physics, McGill University, \\
  3600 rue University,
  Montr\'eal, QC H3A 2T8, Canada}

\emailAdd{renatosa@ift.unesp.br}

\emailAdd{imorrison@physics.mcgill.ca}

\abstract{
  We consider the consequences of global higher-spin symmetries in 
  quantum field theories on a fixed de Sitter background of spacetime 
  dimension $D \ge 3$.
  These symmetries enhance the symmetry group associated with
  the isometries of the de Sitter background and thus strongly
  constrain the dynamics of the theory. In particular,
  we consider the case when a higher spin charge acts linearly on
  a scalar operator to leading order in a Fefferman-Graham expansion
  near the future/past conformal boundaries. We show that this implies
  that the expectation values of the operator inserted near the 
  boundaries are asymptotically Gaussian. 
  Thus, these operators have trivial cosmological spectra,
  and on global de Sitter these operators have only Gaussian correlations
  between operators inserted near future/past infinity.
  The latter result may be interpreted as an analogue of 
  the Coleman-Mandula theorem for QFTs on de Sitter spacetime.
}


\arxivnumber{1511.06753} 

\preprint{NSF-KITP-15-135} 

\begin{document}

\maketitle

\section{Introduction} \label{sec:intro}

Quantum field theories (QFTs) are difficult to solve. 
This statement is already true for QFTs on Minkowski spacetime,
but it becomes emphatically true for QFTs on curved spacetimes.
Quantum field theory on curved backgrounds, including 
perturbative gravity, provides the current framework used to describe
potentially observable quantum effects in cosmology 
\cite{Mukhanov:2005sc}.
Yet very few soluble QFTs exist in this setting, and 
perturbation theory remains the de facto technique to tackling
such theories. The limited range of validity of standard 
perturbative techniques makes it challenging to investigate 
super-Hubble effects, such as those induced by fields with long-range 
order, which might be relevant to eternal inflation and
the observed effective cosmological constant.

QFTs become more tractable when they include
heightened symmetry -- i.e. symmetries not realized in nature -- 
whose presence provides additional handles
with which we may grasp the theories.
A classic example is higher spin (HS) symmetry. HS symmetries are
symmetries whose generators transform has rank $n > 1$ tensors
with respect to the spacetime isometry group.
Typically, the presence of these symmetries in quantum theories is so 
constraining that at least some aspects of a theory may be solved 
exactly.
In $D > 2$ dimensional Minkowski space, the Coleman-Mandula theorem 
\cite{Coleman:1967ad} asserts that HS symmetries constrain the 
scattering matrix of a theory to be trivial. 
In $D=2$ dimensions, the presence of such symmetries 
implies that scattering matrix has no particle production 
(particle number is conserved and scattering is elastic) \cite{Parke:1980ki}.
When combined with conformal symmetry, HS symmetries
yield interesting classes of soluble rational CFTs in 2D
\cite{Zamolodchikov:1985wn}, and in 3D CFTs they
constrain the theory to have a free-field current
algebra, and thus be essentially free \cite{Maldacena:2011jn}.
In each of these cases, the presence of HS symmetries forces
the underlying theory to behave as a free theory for at least a
large set of physical observables.

In this paper we investigate consequences of HS symmetries
in QFTs on a fixed de Sitter (dS) spacetime. de Sitter space provides
the maximally symmetric cosmological model of an inflating spacetime,
and is the lowest-order solution in the standard slow-roll expansion
of inflation. The natural set of QFT observables we investigate are 
the vacuum expectation values of operators inserted near the 
future and past conformal boundaries.
In the context of inflation, observables located near the future
asymptotic boundary of de Sitter correspond to late-time expectation
values which provide the input for cosmological power spectra.
We also consider the correlation functions of observables located near
both the past and future asymptotic boundaries of global de Sitter.
These correlators describe the global dS analogue 
of a scattering experiment.\footnote{Here we mean only that these correlations
measure the transition amplitude for states constructed at asymptotically
early/late times. We will not attempt to establish a rigorous 
notion of asymptotic particle states for global dS. See \cite{Marolf:2012kh}
for a construction of such particle states
and their corresponding S-matrix for \emph{perturbatively} interacting 
QFTs on global de Sitter space.}
Our basic tactic is to analyze how the Ward identities associated 
with HS symmetries constrain the correlation functions of these
observables. 
In many ways, our investigation is similar in spirit to recent 
cosmological ``consistency conditions'' and ``soft theorems'' 
(see, e.g., \cite{
Maldacena:2002vr,Creminelli:2004yq,Cheung:2007sv,Hinterbichler:2012nm,
McFadden:2014nta} and references therein), though we emphasize that
our analysis does not include gravitational back-reaction.

More concretely, our analysis proceeds as follows. For simplicity we
consider the effect of HS symmetries on correlations of a scalar operator
$\phi(x)$. We assume the theory admits a charge $Q_p^{(s)}$ which is 
the spin $s>1$ analogue 
of a translation in a dS Poincar\'e chart. Although this chart provides
a convenient interpretation for $Q_p^{(s)}$, the charge is in fact
well-defined everywhere on dS.
The action of $Q_p^{(s)}$ on $\phi(x)$ is local and may be written
as a sum of local operators $\cO_A(x)$ of the schematic form
\eq{ \label{eq:intro1}
  \Com{Q_p^{(s)}}{\phi(x)} = \sum_A C_A \cO_A(x) .
}
In general, any operator with the correct quantum numbers may appear
on the right-hand side of this expression, making a general analysis
of (\ref{eq:intro1}) intractable. However, if we are interested in 
correlators of $\phi(x)$ near the conformal boundaries of dS
(i.e., near past/future asymptotic infinity), we may expand 
(\ref{eq:intro1}) in a Fefferman-Graham expansion in powers of 
conformal time $\eta$. 
Only those operators which scale with $\eta$
in the same way as $\phi(x)$ will contribute to the commutator
asymptotically.
Thus, for instance, for a scalar with characteristic scaling
$\phi(x) = O(\eta^\Delta)$, $\Delta > 0$, as $\eta \to 0$,
we may truncate the right-hand side to operators which likewise
scale like $O(\eta^\Delta)$ when evaluating the commutator
at asymptotically late times ($\eta \to 0$).

Here we consider the simple case when $\phi(x)$ and its
descendants are the only operators in the theory which scale 
like $O(\eta^\Delta)$ as $\eta \to 0$. In this case the action of
$Q_p^{(s)}$ on $\phi(x)$ becomes \emph{asymptotically linear} in $\phi(x)$ 
near the conformal boundary, i.e. the action takes the form
\eq{
  \Com{Q_p^{(s)}}{\phi(x)} \bigg|_{O(\eta^\Delta)}
  = \mathscr{D}(x) \phi(x) \bigg|_{O(\eta^\Delta)}  ,
}
where $\mathscr{D}(x)$ is a differential operator.
When this occurs, we show that this implies that the leading
$O(\eta^{n\Delta})$ behavior of an $n$-pt correlation function
of $\phi(x)$ is Gaussian, i.e., it is composed of 2-pt correlations.
The same conclusion holds in global dS for correlation functions
in which each operator is placed near one of the asymptotic
boundaries. Thus $\phi(x)$ has trivial cosmological spectra (no
bispectrum, Gaussian trispectrum, etc.), and also has no
``scattering'' in global dS (when measured with respect to 
equivalent initial/final vacua).

The assumption that the action of a HS charge 
is asymptotically linear is clearly very restrictive. 
However, we regard this assumption as the appropriate dS analogue of one 
of the assumptions of the Coleman-Mandula theorem: a symmetry of the S-matrix
is one which maps $n$-particle states to $n$-particle states 
\cite{Coleman:1967ad}.
Said differently, a symmetry-generating charge acts linearly on the field 
redefinition-invariant parts of the asymptotic correlation functions.
In dS QFT, the leading asymptotic behavior of vacuum correlation functions
near the conformal boundaries is, at least in perturbation theory,
field redefinition-invariant and the key input in the perturbative 
de Sitter S-matrix \cite{Marolf:2012kh}. Thus, we regard our result
as providing an analogue of the Coleman-Mandula theorem for dS QFT.

We note in passing that HS symmetries on (asymptotically) de Sitter
spacetimes have been of recent interest due to their roles in
Vasiliev theories of HS gravity \cite{Vasiliev:1990en} as well as
a potential realization of the de Sitter/Conformal Field Theory (dS/CFT)
correspondence \cite{Anninos:2011ui}.
In general these theories include dynamical gravity which we do 
not consider here. However, when solutions to these theories 
may be described as having an exact de Sitter metric, as well as a matter
sector which satisfies the properties stated in \S\ref{sec:prelims} below,
then our results apply.
The HS charges we consider here are more general
than those which appear in this limit of Vasiliev theory,
as the latter are generated by conserved traceless currents, whereas
the conserved currents we consider need not be traceless.

This paper is organized as follows. 
After briefly reviewing necessary background material in \S\ref{sec:prelims},
we analyze general aspects of HS symmetries in dS QFTs in \S\ref{sec:HS}.
As a concrete and simple example, we provide a discussion
of the HS symmetries present in dS complex Klein-Gordon theory
in \S\ref{sec:free}.
In \S\ref{sec:linear} we consider HS charges which act
linearly on a scalar field everywhere on dS. Not surprisingly, 
when this occurs the correlation functions of the scalar field are
constrained to be everywhere Gaussian.
In \S\ref{sec:Alinear} we consider the more general case
of HS charges whose action becomes linear only asymptotically.
We conclude with a brief discussion in \S\ref{sec:disc}.

\section{de Sitter QFTs} \label{sec:prelims} 

We use this preliminary section to establish our notation as
well as review background material relevant to our study.

We consider $D = d+1$ dimensional de Sitter spacetime $dS_D$
with curvature radius $\ell$.
The spacetime is conveniently described as a hyperboloid
in a $D+1$-dimensional Minkowski embedding space:
\eq{\label{eq:dSsurface}
  dS_{D} := \left\{ X \in \Reals^{D,1} \; | \;
    X \cdot X = \ell^2
  \right\} .
}
This surface is preserved under the action of the embedding space
Lorentz group $SO(D,1)$, i.e. boosts and rotations in the embedding
space which preserve the origin. This group is thus the isometry 
group of $dS_D$ (the ``dS group'').
The entire manifold may be covered by the global coordinate chart
\eq{
  ds^2 = -d\tau^2 + \ell^2 \cosh^2(\tau/\ell) d\Omega_d^2  ,
}
where $d\Omega_d^2$ is the line element of the unit sphere $S^d$.
This chart nicely displays the hyperboloid geometry of dS, and
in particular the fact that the manifold may be foliated by 
compact Cauchy surfaces. The conformal boundary of the spacetime is composed
of two disconnected components, future (past) conformal infinity 
$\mathscr{I}^{+(-)}$ located at $\tau \to +(-)\infty$,
each conformal to $S^d$.

For our purposes, a more convenient set of coordinates is 
given by the (expanding) Poincar\'e chart:
\eq{ \label{eq:gPoincare}
  ds^2 = \frac{\ell^2}{\eta^2}
  \left( -d \eta^2 + \delta_{ab}dx^a dx^b \right) ,
  \quad \eta \in (-\infty,0) , \; x^a \in \Reals^d .
}
Here $\eta$ is conformal time, 
roman indices run over spatial dimensions $1,\dots,D$,
and $\delta_{ab}$ is the flat metric on $\Reals^d$.
The expanding Poincar\'e chart covers only half of the dS manifold;
the other half manifold may covered by a contracting Poincar\'e
chart with conformal time $\eta \in (0,+\infty)$.
In these coordinates the dS isometries may be described as:
(i) translations and (ii) rotations on constant-$\eta$ surfaces,
(iii) the dilations $x^\mu \to \l x^\mu$, and (iv) special conformal
transformations $x^\mu \to x^\mu+2 b^\nu x_\nu x^\mu - x_\nu x^\nu b^\mu$
generated by vectors $b^\mu$ tangent to constant-$\eta$ surfaces.

With regard to any compact subset of spacetime,
in the limit $\ell \to \infty$ the de
Sitter geometry reduces to Minkowski spacetime.
We refer to this limit, with all other dimensionful quantities
held fixed, as the flat-space limit.
The Poincar\'e coordinates (\ref{eq:gPoincare}) are not well-suited
for this limit; instead  we use the ``proper time'' coordinate $t$:
\eq{
  \eta = -\ell e^{-t/\ell}, \quad t = -\ell \ln\left(-\frac{\eta}{\ell}\right),
}
so that the line element becomes
\eq{
  \label{eq:gproper}
  ds^2 = -dt^2 + e^{2t/\ell} \delta_{ab}dx^a dx^b , \quad t \in \Reals .
}
Taking $\ell \to \infty$ with these coordinates fixed, the line element
indeed reduces to that of Minkowski space.

Given two points $X_i, X_j \in dS_D$ it is convenient to define
the $SO(D,1)$-invariant \emph{chordal distance}\footnote{This is actually
one quarter the chordal distance.}
\eq{
  X_{ij} := \frac{1-X_i\cdot X_j/\ell^2}{2} .
}
Different causal relationships between the points are encoded in 
$X_{ij}$ as follows:\footnote{Spacelike-separated points in $dS$
which may be connected by a geodesic satisfy $X_{ij} \in (0,1]$.}
\eqn{
  \text{spacelike separation:}& & \quad X_{ij} > 0 ,
  \nn \\
  \text{null separation:}& & \quad X_{ij} = 0,
  \nn \\
  \text{timelike separation:}& & \quad X_{ij} < 0 .
}
The chordal distance may be expressed in Poincar\'e coordinates as
\eq{
  X_{ij} = \frac{|\vx_i -\vx_j|^2 -(\eta_i-\eta_j)^2}{4\eta_i \eta_j}  .
}

We are interested in studying local dS QFTs
which have standard properties of QFTs on curved spacetime
(see e.g., \cite{Hollands:2008vx,Hollands:2014eia},
as well as discussion in \cite{Wald:1995yp}).
In particular, we restrict our attention to theories with
the following properties:

\begin{enumerate}[i)]

\item {\bf dS covariance:} \label{i:covariance}
  The theory does not select a preferred direction
  or otherwise spoil the unitary representation of the 
  dS isometry group.
  For each dS Killing vector field (KVF) $\xi^\mu$
  there exists an isometry generator in the QFT of the form
  \eq{
    G_\xi := \int d\Sigma(x) n^\mu \xi^\nu T_{\mu\nu}(x) \bigg|_\Sigma ,
  }
  where $\Sigma$ is a Cauchy surface, $n^\mu$ is
  the future-pointing unit normal vector, and 
  $T_{\mu\nu}(x)$ is the QFT stress-tensor. The generators satisfy
  the $SO(D,1)$ algebra inherited from the KVFs
  \eq{
    \Com{G_{\xi_1}}{G_{\xi_2}} = - i G_{\Com{\xi_1}{\xi_2}} ,
  }
  where for vector fields $\Com{\xi_1}{\xi_2}$ denotes the Lie bracket.
  The generators act on any quantum operator $\cO(x)$ via a Lie derivative:
  \eq{
    \Com{G_\xi}{\cO(x)} = i \Lie_\xi \cO(x) .
  }

\item {\bf dS-invariant states:} \label{i:dSIstates}
  The theory admits at least one state invariant under the action
  of the dS isometry group $SO(D,1)$.\footnote{%
    Not all theories satisfying (\ref{i:covariance}) necessarily
    have such states. A well-known example is the massless, minimally-coupled
    Klein-Gordon field, when quantized via canonical 
    quantization \cite{Mottola:1984ar,Allen:1985ux}.}

\item {\bf microlocal spectrum condition (``$\mu$SC''):} \label{i:MSC}
  Correlation functions of the theory contain short-distance (ultraviolet)
  singularities consistent with the micro-local spectrum condition
  of \cite{Brunetti:1995rf}.
  Essentially, the $\mu$SC states that the only singularities of 
  correlation functions are at coincident configurations, and are of 
  ``positive frequency'' as measured in any locally flat coordinate chart. 
  This assures that the singularities present in correlation functions 
  look like those of the usual Minkowski vacuum.
  The $\mu$SC may be regarded as the generalization of
  the Hadamard condition to interacting theories
  \cite{Radzikowski:1996aa,Brunetti:1995rf,Brunetti:1999jn}.

\item {\bf IR-regularity:} \label{i:IR}
  We assume that correlation functions of local operators do not
  grow too quickly as the chordal distance $Z$ between two clusters of
  operators grows.
  For massive theories in dS, such correlation functions \emph{decay} 
  as the chordal separation $Z$ increases 
  \cite{Marolf:2010nz,Hollands:2010pr,Korai:2012fi}.
  For interacting massless theories, such correlation functions are
  known to grow in perturbation theory like a power of ($\log Z$)
  (see \cite{Hollands:2011we,Rajaraman:2010xd} for discussion of
  dS-invariant states, as well as 
  \cite{Miao:2010vs,Prokopec:2003bx,Kitamoto:2011yx,Burgess:2010dd}
  for less-symmetric states).
  To be concrete, we assume that the correlation functions of local
  operators clustered into two groups separated by chordal distance
  $|Z| \gg 1$ may be bounded by $c Z$ for some finite constant 
  $c$.\footnote{This assumption seems eminently reasonable for
  any theory which might called massive or massless. However, in
  de Sitter there also exist unitary, tachyonic theories which
  violate this assumption. The simplest example is a discrete set of
  free, tachyonic scalars \cite{Bros:2010aa}. The correlation functions of these
  theories can grow like a power of $Z$. However, we know of no
  interacting theories which violate assumption (\ref{i:IR}).}

\end{enumerate}

Given these assumptions, we expect that correlation functions of 
scalar operators with respect to a dS-invariant state $\Omega$ admit
representations as generalized Mellin-Barnes integrals
which take the form \cite{Marolf:2010nz,Marolf:2011aa,Hollands:2010pr,
  Hollands:2011we,Korai:2012fi}:
\eqn{ \label{eq:MBform}
  \C{\phi(x_1)\dots\phi(x_n)}_\Omega
  = \left[\prod_{i<j}^n \int_{C_{ij}} 
    \frac{d\mu_{ij}}{2\pi i} X_{ij}^{\mu_{ij}}\right] \cM(\vec{\mu}) .
}
There is one integration variable $\mu_{ij}$ for each pairing $x_i, x_j$. 
Each $\mu_{ij}$ is integrated along a Mellin-Barnes contour $C_{ij}$, i.e. a 
contour traversed from $-i\infty$ to $+i\infty$ in the left half-plane,
which may be diverted to left or right to avoid pole singularities in 
$\cM(\vec{\mu})$.
We refer to $\cM(\vec\mu) := \cM(\mu_{12},\dots,\mu_{n-1,n})$ 
as the \emph{Mellin amplitude} of the correlation
function $\C{\phi(x_1)\dots\phi(x_n)}_\Omega$.
Mellin amplitudes contain pole singularities in the complex $\mu_{ij}$ planes.
The contour integrals converge due to the fact that $\cM(\vec{\mu})$
decays sufficiently fast -- i.e. exponentially -- 
as any $|\Im \mu_{ij}| \to \infty$. For the generalized Mellin transforms
to be defined at timelike separations one must add the standard
position-space $i\epsilon$ prescriptions to the $X_{ij}$. These 
$i\epsilon$ prescriptions will be unimportant to our analysis so we
will suppress them.

The details of these Mellin-Barnes representations will not be
important to our analysis. The key point to take away is that these
representations define functions of the chordal distances regarded
as independent variables. The set of chordal distances which 
describe $n$ points in dS are not all independent as they
are constrained so that all $n$ points ``fit'' in $dS_D$. 
However, the Mellin-Barnes integrals (\ref{eq:MBform}) define 
functions of the $X_{ij}$ over a larger domain.
Further details of the Mellin-Barnes representation are presented
in \cite{Marolf:2010nz,Hollands:2010pr,Hollands:2011we}.\footnote{%
  Mellin transforms are a natural integral transform to consider 
  whenever the underlying function has power-law asymptotics,
  As such, Mellin transforms are also useful tools in
  CFT and AdS/CFT (see, e.g.,
  \cite{Fitzpatrick:2011ia,SimmonsDuffin:2012uy} and references therein).
  Historically, Mellin transforms have played a key role in calculations
  of multi-loop Feynman diagrams in Minkowski space \cite{Smirnov:2004ym}).}

\section{Higher spin symmetries on dS} \label{sec:HS} 

The higher-spin symmetries we study are generated by a local, covariantly
conserved current $J_{\mu_1\dots\mu_s}(x)$ of rank $s>2$.
Given a rank-$(s+1)$ current and a rank $s$ Killing tensor
$K^{\nu_1\dots\nu_s}(x)$ we may define a spin $s$ conserved charge\footnote{%
  We do not require the current $J_{\mu\nu_1\dots\nu_s}(x)$ 
  nor the Klling tensor $K^{\nu_1\dots\nu_s}(x)$ be traceless.}
\eq{
  Q^{(s)}_K := \int d\Sigma(x)\, n^\mu K^{\nu_1\dots\nu_s}
  J_{\mu\nu_1\dots\nu_s}(x)  \Big|_\Sigma .
}
Here $\Sigma$ is a \emph{global} dS Cauchy surface and
$n^\mu$ is the future-pointing normal vector. 
Any de Sitter Killing tensor may be
written as a product of KVFs \cite{Thompson:1986aa},
KVFs, i.e.,
\eq{ \label{eq:Ks}
  K^{\nu_1\dots\nu_s}(x) = \xi_1^{\nu_1}\cdots\xi_s^{\nu_s}(x) ,
}
where each $\xi_i^\mu$ may be a distinct KVF.
Using this form it is easy to show that the charges
$Q_K^{(s)}$ and the dS generators $G_\xi$ have the commutators
\eq{ \label{eq:GQ}
  \Com{G_\xi}{Q_K^{(s)}} = - i Q^{(s)}_{\Lie_\xi K} .
}
Since $\xi$ is a dS KVF, the action of the lie derivative $\Lie_\xi$
on $K^{\nu_1\dots\nu_s}(x)$ produces another Killing tensor of the form 
(\ref{eq:Ks}).
Thus, the charges $Q^{(s)}_K$ and the dS generators $G_\xi$ form
a closed algebra which enlarges the $SO(D,1)$ algebra of the 
generators alone.

We will be primarily concerned with the higher-spin charges
\eq{
  Q_p^{(s)} := \int d\Sigma(x)\, n^\mu p^{\nu_1}\dots p^{\nu_s}
  J_{\mu\nu_1\dots\nu_s}(x)  \Big|_\Sigma ,
}
where $p^\mu$ is a KVF whose flow corresponds to
translation along constant $\eta$ surfaces in some Poincar\'e chart.
Note that for a given $p^\mu$ the Poincar\'e time direction $\d_\eta$ is 
unique up to choice of sign (the sign corresponds to whether the 
Poincar\'e chart is expanding or contracting).
It is convenient to adopt this expanding Poincar\'e chart to 
describe characteristics of $Q_p^{(s)}$.
We emphasize, however, that $Q_p^{(s)}$ is defined over the
global dS manifold.
We normalize $p^\mu$ such that in this Poincar\'e 
chart $\delta_{ab} p^a p^b = 1$. 
Let us denote the dS generators in this chart by $P_a$ (translations),
$D$ (dilations), $R_{ab}$ (rotations), and $K_a$ (SCTs).
Then, for instance, $Q^{(1)}_p$ corresponds to a linear combination 
of the $P_a$'s.
From (\ref{eq:GQ}) it follows that the $Q_p^{(s)}$ enjoy 
simple commutation relations with some of the dS generators:
\eq{ \label{eq:QpCRs}
  \Com{P_a}{Q_p^{(s)}} = 0, \quad
  \Com{D}{Q_p^{(s)}} = -s Q_p^{(s)} ,\quad
  \Com{R^\perp}{Q_p^{(s)}} = 0 .
}
Here $R^\perp$ represents the generators of rotations which preserves 
$p^\mu$ (these exist for $D \ge 4$).
In general $Q_p^{(s)}$ are covariant under $SO(d)$ rotations in the
sense that
\eq{
  R(\alpha) Q_p^{(s)} R^{-1}(\alpha) = Q^{(s)}_{R^{-1}p} .
}
On the other hand, the special conformal transformation generators
$K_a$ alter $Q_p^{(s)}$ in a complicated way, and $\Com{K_a}{Q_p^{(s)}}$
does not correspond to any $Q_p^{(s)}$.

The action of a HS charge $Q_p^{(s)}$ on a local operator $\cO(x)$
is given by the commutator $\Com{Q_p^{(s)}}{\cO(x)}$.
For simplicity we will focus on the action of
$Q_p^{(s)}$ on scalar operators. The most general action is 
given by the following:
\begin{lemma} \label{lemma:Qphi}
  Let $\phi(x)$ be a local scalar operator. The most general form
  of the commutator $\Com{Q_p^{(s)}}{ \phi(x)}$ is 
  \eq{ \label{eq:Qphi}
    \Com{Q_p^{(s)}}{ \phi(x)} = \sum_A \frac{1}{\eta^{s-k}}
    C_A^{\mu_1\dots\mu_k} \cO_{\mu_1\dots\mu_k}^A(x) .
  }
  Here $A$ is a collective index labeling operators $\cO^A_{\mu_1\dots\mu_k}(x)$
  which transform covariantly under $SO(D,1)$,
  $k$ is an integer (which depends on $A$) satisfying $0 \le k \le s$, 
  and the $C_A^{\mu_1\dots\mu_k}$ are constant coefficients. These
  coefficients are composed
  of products of $p^\mu$ and $t^\mu$, where $t^\mu \d_\mu = \d_\eta$, such 
  that there is an even (odd) number of $p^\mu$'s when $s$ is even (odd).
\end{lemma}
We prove this lemma in Appendix~\ref{app:lemmas}. 
Note that in order for the dimensions to be consistent in (\ref{eq:Qphi}), 
the operators $\cO^A_{\mu_1\dots\mu_k}(x)$ must have length dimension
$1-D/2-k$.
The explicit factors of $\eta$ and $t^\mu$ are allowed in (\ref{eq:Qphi}) 
because $p^\mu$ selects a unique time coordinate $\eta$.

%
%
Let us compare lemma~\ref{eq:Qphi} to the 
analogous result in Minkowski QFT. 
If a Minkowski theory has a HS current then then one may 
construct, e.g., the HS charge
\eq{
  Q_1^{(s)} := \int d^dx J_{01\dots1}(x) \bigg|_{x^0={\rm const}} ,
}
where we use standard Cartesian coordinates $\{x^0,x^1,\dots,x^d\}$.
For $s>1$ this is the higher spin analogue of a translation along $x^1$. 
It is easy to show that in this case the action of $Q_1^{(s)}$
on a scalar field is of the form
\eq{ \label{eq:MinkowskiQphi}
  \Com{Q_1^{(s)}}{\phi(x)} = \sum_A c_A \cO^A_{1\dots1}(x) ,
}
where $c_A$ are constant coefficients. Comparing (\ref{eq:MinkowskiQphi})
to (\ref{eq:Qphi}), we see that if a dS theory is to admit a smooth
flatspace limit, it must be that 
terms involving $t^\mu$ in (\ref{eq:Qphi}) must vanish as
$\ell \to \infty$, either due to explicit factors of $1/\eta$ (which
tend to zero in the limit), or because the operator vanishes in the limit.

Returning to the dS context,
we assume that there exist dS-invariant states which are annihilated by 
$Q_{p}^{(s)}$. It follows that expectation
values taken with respect to these states satisfy ``charge conservation
identities,'' or Ward identities, obtained by commuting $Q_p^{(s)}$
through the string of operators. E.g., for such a state $\Omega$
we may commute $Q_p^{(s)}$ through $\C{\phi(x_1)\phi(x_2)\dots\phi(x_n)}_\Omega$
to obtain
\eq{ \label{eq:WardId}
  0 = \C{\Com{Q_p^{(s)}}{\phi(x_1)}\phi(x_2)\dots\phi(x_n)
    + \dots 
    + \phi(x_1)\dots\phi(x_{n-1})\Com{Q_p^{(s)}}{\phi(x_n)}}_\Omega .
}
These Ward identities will be the central object of our study.

\section{HS charges in free fields} \label{sec:free} 

In order to provide a concrete example of HS symmetries,
in this section we review the HS symmetries present in complex 
Klein-Gordon theory on dS.

A massive, complex Klein-Gordon field on dS may be described by
the classical action
\eq{ \label{eq:Sfree}
  S = \int d^D x \sqrt{-g} \left( - \nabla_\mu \phi^{\dagger} \nabla^\mu \phi(x)
     - M^2 \phi^\dagger\phi(x) \right),
   \quad M^2 > 0 .
}
In general we let $M^2 = m^2 + \xi R(x)$, with $m^2 > 0$, $\xi$
a coupling constant, and $R(x)$ the Ricci scalar which is constant
on dS. The case $m^2 = 0$, $\xi = (d-1)/(4d)$ corresponds to a 
conformally-invariant theory; in terms of $M^2$
this ``conformally coupled'' mass is
\eq{
  M_{\rm c.c}^2\ell^2 = \frac{d^2-1}{4} .
}
Upon quantization
the theory possesses a unique dS-invariant state $\Omega$ satisfying the 
$\mu$MC \cite{Mottola:1984ar,Allen:1985ux}. 
The composite operators below are defined by normal ordering with respect to 
$\Omega$.

The reader is undoubtedly familiar with the lowest-spin currents in the theory, namely the spin-1 ``Klein-Gordon current'' and the stress tensor:
\eqn{
  \label{eq:J1free}
  J_\mu(x) &=& \phi^{\dagger} \overleftrightarrow{\nabla_\mu} \phi(x) , \\
  \label{eq:J2free}
  J_{\mu\nu}(x) &=& 2\nabla_{(\mu} \phi^{\dagger} \nabla_{\nu)} \phi(x)
  + 2 \xi \nabla_\mu \nabla_\nu \left( \phi^\dagger \phi(x)\right)
  - 2 \xi R_{\mu\nu} \phi^\dagger \phi(x)
  \nn \\ & &
  - g_{\mu\nu}\left( 
    \nabla_\l \phi^{\dagger} \nabla^\l \phi(x)
    + 2\xi \Box \left(\phi^\dagger \phi(x)\right)
    + M^2 \phi^\dagger\phi(x) \right),
}
where $R_{\mu\nu} = (d/\ell^2)g_{\mu\nu}$ is the dS Ricci tensor.
Perhaps less familiar is the fact that the theory admits symmetric,
covariantly conserved currents of every rank which are of the form
\eq{
  J_{\mu_1\dots\mu_n}(x)
  = \sum_{j=0}^n c_j \nabla_{(\mu_1} \dots \nabla_{\mu_j} \phi^{\dagger}
  \nabla_{\mu_{j+1}}\dots \nabla_{\mu_n)}\phi(x) + {\rm traces} ,
}
where the $c_j$ are constants and ``traces'' denote terms composed of 
partial traces of the terms written, multiplied by appropriate factors of 
the metric.\footnote{
  In the classical field theory, the currents may also be made traceless 
  when $M^2 = M^2_{\rm c.c.}$. However, as in the familiar case of the 
  stress tensor, we expect that this tracelessness may be spoiled in
  the quantum theory by anomalies due to the curved background.
  We thank E.~Mottola for raising this point. 
}
The most tidy example is the spin-3 current, which with convenient 
normalization may be written
\eqn{
  \label{eq:J3free}
  J_{\mu\nu\l}(x) &=& \frac{1}{4(d+2)} \bigg[
  (d-1) \left( \phi^{\dagger} \nabla_{(\mu} \nabla_\nu \nabla_{\l)} \phi(x)
    - \nabla_{(\mu} \nabla_\nu \nabla_{\l)}  \phi^{\dagger} \phi(x)
  \right)
  \nn \\ & & \ph{\frac{1}{4(d+2)} \bigg[}
  -3(3+d)  \nabla_{(\mu} \phi^{\dagger}\overleftrightarrow{\nabla_\nu}\nabla_{\l)} \phi(x)
  + 6 g_{(\mu\nu}
  \nabla^\a \phi^{\dagger} \overleftrightarrow{\nabla_\l} \nabla_{\a)}\phi(x)
  \nn \\ & & \ph{\frac{1}{4(d+2)} \bigg[}
  +  \left[6 M^2 - (d-1)(3d+2)\ell^{-2}\right]
  g_{(\mu\nu} J_{\l)}(x) \bigg] ,
}
and which has trace
\eq{
  g^{\mu\nu} J_{\mu\nu\l}(x)
  = \left(M^2 - M_{\rm c.c.}^2 \right) J_\l(x) .
}
Unfortunately, we are unaware of explicit expressions for these
HS currents for general $M^2$; for the conformally coupled case 
expressions for the currents may be obtained from known
CFT results (see, e.g., \cite{Bakas:1990ry,Mikhailov:2002bp}).

Let us examine the action of the resulting HS charges on $\phi(x)$.
A straightforward if tedious way to compute
the commutator $\Com{Q_p^{(s)}}{\phi(x)}$ is by direct application of
the canonical commutation relations. Expressed at equal times 
in Poincar\'e coordinates, these familiar relations are
\eq{
  \Com{\phi(\eta,\vx)}{\pi(\eta,\vy)} 
  = i \left(\frac{\eta}{\ell}\right)^d \delta^d(\vx,\vy) ,
}
where $\pi(x)$ is the momentum conjugate to $\phi(x)$ and
$\delta^d(\vx,\vy)$ is the $d$-dimensional Dirac delta function.
For example, for the spin-2 charge associated with the spin-3 current
(\ref{eq:J3free}), diligent calculation yields
\eq{
  \Com{Q_p^{(2)}}{\phi(x)} = - i \frac{|\vec{p}|^2}{2(d+2)}
  \left(
    -\d_\eta^2 + \frac{(d-1)}{\eta} \d_\eta - \frac{M^2\ell^2}{\eta^2}
    + \Delta_s \right) \phi(x)
    + i \d_p^2 \phi(x) .
}
Here $|\vec{p}|^2 = \delta_{ab} p^a p^b$, $\d_p = p^\mu \d_\mu$ is set
to unity in the main text, and $\Delta_s$ is the Laplacian
compatible with the flat metric on constant-$\eta$ hypersurfaces.
The terms in parenthesis are proportional to the KG wave operator
and thus annihilate the field $\phi(x)$ on-shell. The final
expression is then
\eq{
  \Com{Q_p^{(2)}}{\phi(x)} =  i \d_p^2 \phi(x) .
}
For general $s$ it is more efficient to use lemma~\ref{lemma:Qphi} in
order to prove that the commutator is
\eq{ \label{eq:QphiFree}
  \Com{Q^{(s)}_p}{\phi(x)} = i \d_p^s \phi(x) .
}
The argument is as follows. Since the currents are bi-linear in $\phi(x)$,
and since the canonical commutation relations map
$\phi \times \phi \to \mathbb{C}$, it follows that the right-hand side of 
(\ref{eq:QphiFree}) must be linear in $\phi(x)$. The commutator 
is a solution to the Klein-Gordon equation, the thus right-hand side
of (\ref{eq:QphiFree}) must also be a solution. 
The only term which is linear in $\phi(x)$,
a solution to the Klein-Gordon equation, and consistent with 
lemma~\ref{lemma:Qphi} is $\d_p^s \phi(x)$.

\section{HS charges with linear action} \label{sec:linear} 

In this section we consider HS charges which act linearly on a 
scalar field $\phi(x)$. By this we mean that 
\eq{ \label{eq:linearAction}
  \Com{Q_p^{(s)}}{\phi(x)} = \mathscr{D}(x) \phi(x) ,
}
where $\mathscr{D}(x)$ is a differential operator of the form
\eq{ \label{eq:LinearD}
  \mathscr{D}(x) = \sum_A \frac{1}{\eta^{s-k} }
  C_A^{\mu_1\dots\mu_k} \mathscr{D}^{(A)}_{\mu_1\dots\mu_k} ,
}
where the coefficients $C_A^{\mu_1\dots\mu_k}$ are as in lemma~\ref{lemma:Qphi},
and the $\mathscr{D}^{(A)}_{\mu_1\dots\mu_k}$ are rank-$k$ covariant
differential operators composed of products of 
$\nabla_\mu$, $g_{\mu\nu}$, and $\Box$. Within this set-up we shall
prove the following:

\begin{lemma}\label{lemma:linear}
  Consider a QFT satisfying the properties of \S\ref{sec:prelims}
  in spacetime dimension $D \ge 3$.
  Let $\Omega$ be a dS-invariant state which is annihilated by the 
  HS charge $Q^{(s)}_p$. Suppose that
  the action of $Q^{(s)}_p$ on a scalar field $\phi(x)$ is linear,
  and furthermore that $\mathscr{D}(x)$ contains the term 
  $(p^\mu \d_\mu)^s$. Then the correlation functions
  $\C{\phi(x_1)\dots\phi(x_n)}_{\Omega}$ are Gaussian.
\end{lemma}

\proof
Consider the Ward identity associated with commuting $Q_p^{(s)}$ through
the correlation function 
\eq{
  F := \C{\phi(x_1)\dots\phi(x_n)}_\Omega ,
}
where no pair of points is null-separated.
We may regard $F$ as a function of the $n(n-1)/2$ chordal 
distances $X_{ij}$.
Due to the linear action (\ref{eq:linearAction}) of $Q_p^{(s)}$, this
Ward identity may be written as
\eq{ \label{eq:linearWard}
  0 =  \sum_{k=1}^n \mathscr{D}(x_k) F .
}
Unpacking this expression results in several terms. Let us focus
on terms generated by $[p^\mu (\d/\d x_1^\mu)]^s$.
From the derivatives
\eqn{
  \left(p^\mu \frac{\d}{\d x_1^\mu} \right) X_{12}
  &=& \frac{\vec{p}\cdot (\vec{x}_1-\vec{x}_2)}{2\eta_1\eta_2}
  = - \left(p^\mu \frac{\d}{\d x_2^\mu} \right) X_{12} , 
  \nn \\
  \left(p^\mu \frac{\d}{\d x_1^\mu}\right)^2 X_{12}
  &=& \frac{p^2}{2\eta_1 \eta_2}
  =  \left(p^\mu \frac{\d}{\d x_2^\mu}\right)^2 X_{12} ,
}
it follows that that (\ref{eq:linearWard}) contains the terms
\eqn{ \label{eq:T}
  T_k &:=& 
  \left( \frac{1}{2\eta_1\eta_2} \vec{p}\cdot (\vx_1-\vx_2) \right)^{s-k}
  \left( \frac{1}{2\eta_1\eta_3} \vec{p}\cdot (\vx_1-\vx_3) \right)^{k}
  \left(\frac{\d}{\d X_{12}}\right)^{s-k} 
  \left(\frac{\d}{\d X_{13}}\right)^{k} F ,
  \nn \\ & & {\rm for}
  \quad k = 1,\dots,s-1.
}
Each $T_k$ depends on 
$\vec{p}\cdot(\vec{x}_1-\vec{x}_2)$ and 
$\vec{p}\cdot(\vec{x}_1-\vec{x}_3)$ in a distinct way which cannot 
arise from any other term in the Ward identity. 
In particular, these terms can
only arise from the derivative operator $(p^\mu \d/\d x_1^\mu)^s$, and 
can only arise from one way of distributing the derivatives 
$\d/\d x^\mu_1$.\footnote{The case $D=2$ is different.
In this case there is only one spatial dimension, so $p^\mu \d_\mu$
does not have the same effect of selecting a preferred spatial direction.
One may produce terms with the same coordinate dependence of $T_k$ 
by acting with time derivatives $\d_{\eta_1}$.
As a result, it is no longer the case that the $T_k$ must vanish.}
Thus, each $T_k$ must vanish individually in order for the Ward identity to 
be satisfied.
The first line of (\ref{eq:T}) does not vanish for general configurations
of points, and thus the factor on the second line must vanish.
When $n > 3$ we may repeat this argument for terms of the form
$T_k$ but with $x_2$, $x_3$ replaced with other combinations of
points in $\{ x_2,\dots x_n\}$. Ultimately we conclude that $F$ must
satisfy
\eq{ \label{eq:Fconstraint}
  \left(\frac{\d}{\d X_{1i}}\right)^{s-k} 
  \left(\frac{\d}{\d X_{1j}}\right)^{k} F = 0, 
  \quad i,j \in \{2,\dots,n\} ,
  \quad k = 1,\dots,s-1 .
}

We distinguish two ways the equalities (\ref{eq:Fconstraint}) may be satisfied: 
i) $F$ depends only on one chordal distance $X_{1i}$, or
ii) $F$ depends on more than one chordal distance $X_{1i}$, but must
depend on each distance polynomially.
In fact, the most general $F$ is a sum of terms,
each of which satisfies either (i) or (ii).
However, possibility (ii) violates our assumption of IR regularity
((\ref{i:IR}) in \S\ref{sec:prelims}).
If $F$ depends polynomially on all chordal distances
involving $x_1$, then $F$ grows polynomially in the chordal distance
as $x_1$ is taken to large mutual chordal separation from the remaining points.
Thus we conclude that $F$ must be a sum of terms, each of which depends on
only one chordal distance $X_{1 i}$.

We can now repeat the argument for those terms in the Ward identity 
which are generated by $(p^\mu \d/\d x_j^\mu)^s$, $j=2,\dots,n$, and 
which yield constraints similar to (\ref{eq:Fconstraint}) but with
$x_1$ swapped for another point. 
Eventually we are led to conclude that $F$ is a sum of terms, each
of which depends on only one chordal distance per spacetime point.
Thus $F$ is Gaussian. This concludes the proof. $\blacksquare$

\section{HS charges with asymptotically linear action} 
\label{sec:Alinear} 

Next we consider HS charges with a less restrictive
form of action on scalar operators. Here we consider actions
which become linear only in the neighborhood of the asymptotic
boundaries. As operator expressions, the commutator 
$\Com{Q_p^{(s)}}{\phi(x)}$ and Ward identity may be
expanded as a Laurent expansion with respect to conformal time $\eta$
(or the coordinate $q$ defined below) 
in the spirit of a Fefferman-Graham expansion.
For simplicity we focus on the case where $\phi(x)$ is a
scalar operator with characteristic leading behavior near the
conformal boundary
\eq{
  \phi(x) = O(\eta^\Delta), \quad \Delta >  0, \quad {\rm as}\; \eta \to 0.
}
Scalars with this asymptotic form, and with $0 < \Delta < d/2$, 
are often referred to as ``light'' fields, because in the canonical 
example of a Klein-Gordon theory such
operators correspond to fields with mass of order $\ell^{-2}$.
Our results below are valid for any positive $\Delta$.

\begin{definition}\label{def:AL}
  Let $\phi(x)$ be a scalar operator on $dS_D$
  with characteristic scaling
  $\phi(x) = O(\eta^\Delta)$, $\Delta >0$, as $\eta \to 0$. 
  Then the action of charge $Q$ on $\phi(x)$ 
  is \emph{asymptotically linear} if the leading term in the 
  commutator takes the form
  \eq{
    \Com{Q}{\phi(x)} \Big|_{O(\eta^\Delta)} 
    = \mathscr{D}(x) \phi(x)) \Big|_{O(\eta^\Delta)} ,
  }
  where $\mathscr{D}(x)$ is a differential operator of the form
  described in lemma~\ref{lemma:linear}.
\end{definition}

\begin{theorem}\label{thm:Alinear1}
  Let $\phi(x)$ be a scalar operator on $dS_D$, $D \ge 3$,
  with characteristic scaling
  $\phi(x) = O(\eta^\Delta)$, $\Delta > 0$, as $\eta \to 0$, 
  and let $\Omega$ be a 
  dS-invariant state annihilated by the HS charge $Q_p^{(s)}$.
  If the action of $Q_p^{(s)}$ on $\phi(x)$ is asymptotically
  linear and contains the term $(p^\mu \d_\mu)^s$,
  then the leading $O(\eta^{n \Delta})$ behavior of the 
  equal-time correlation functions
  $\C{\phi(\eta,\vec{x}_1)\dots\phi(\eta,\vec{x}_n)}_\Omega$
  is Gaussian.
\end{theorem}

\proof
After a bit of simplification the proof is very similar to 
that of lemma~\ref{lemma:linear}.
Let $F = \C{\phi(\eta,\vec{x}_1)\dots\phi(\eta,\vec{x}_n)}_\Omega$
be an equal-time correlation function evaluated at
$n$ non-coincident points.
The Ward identity now implies that as $\eta \to 0$
\eq{
  \sum_{k=1}^n \mathscr{D}(x_k) F = O(\eta^{n\Delta+\epsilon}) ,
  \quad \epsilon > 0 .
}
Once again we focus on the
terms arising from the derivative $(p^\mu \d/\d x_1^\mu)^s$
within $\mathscr{D}(x_1)$. This yields terms of the form $T_k$ as in 
(\ref{eq:T}). Due to their unique dependence on 
$\vec{p} \cdot (\vec{x}_1 - \vec{x_2})$ and 
$\vec{p} \cdot (\vec{x}_1 - \vec{x_3})$, each of these
terms must satisfy the fall-off condition individually, i.e. each
must be $O(\eta^{n \Delta+\epsilon})$.
Following through the same logic as in the previous proof, we quickly 
conclude that
\eq{ 
  \label{eq:Fconstraint2}
  \left(\frac{\d}{\d X_{1i}}\right)^{s-k} 
  \left(\frac{\d}{\d X_{1j}}\right)^{k} F = O(\eta^{n \Delta+\epsilon}), 
  \quad i,j \in \{2,\dots,n\} ,
  \quad k = 1,\dots,s-1 .
}

To proceed further let $x_{ij} := |\vec{x}_i - \vec{x}_j|^2/4$ so that
\eq{
  X_{ij} = \frac{x_{ij}}{\eta^2} \; ;
}
then (\ref{eq:Fconstraint2}) may be written 
\eq{ \label{eq:Fconstraint3}
  \left(\frac{\d}{\d x_{1i}}\right)^{s-k} 
  \left(\frac{\d}{\d x_{1j}}\right)^{k} F = O(\eta^{n \Delta-2s+\epsilon}), 
  \quad i,j \in \{2,\dots,n\} ,
  \quad k = 1,\dots,s-1 .
}
We next expand $F$ in a Laurent expansion with respect to $\eta$. 
The leading term is
\eq{
  F =
  \C{\phi(\eta,\vec{x}_1) \dots \phi(\eta,\vec{x}_n)}_\Omega 
  \bigg|_{O(\eta^{n\Delta})} =: \eta^{n\Delta} f ,
}
where $f$ is a function of the $x_{ij}$ but does not depend on $\eta$.
We may write $f$ explicitly in terms of Mellin amplitude $\cM(\vec{\mu})$
of $F$ as
\eq{ \label{eq:fcontour}
  f :=
  \left[ \prod_{i < j}^n \int_{C_{ij}}^{'} \frac{d\mu_{ij}}{2\pi i} 
    x_{ij}^{\mu_{ij}}  \right]
  \mathcal{M}(\vec{\mu}) ,
}
where the prime on the integrals denotes that contours are traversed
such that $(\sum_{i <j}^n \mu_{ij} + \frac{n \Delta}{2}) = 0$.\footnote{%
  There will be singularities in $\cM(\vec{\mu})$ at points along
  this set of contours, so (\ref{eq:fcontour}) includes both residue
  and principal parts.
}
The key point is that $f$ may be regarded as a function 
of independent variables $x_{ij}$. It follows that $f$ satisfies 
\eq{ \label{eq:Fconstraint3}
  \left(\frac{\d}{\d x_{1i}} \right)^{s-k} 
  \left( \frac{\d}{\d x_{1j}} \right)^k f = 0, \quad
  k = 1,\dots,s-1 .
}

The equation (\ref{eq:Fconstraint3}) is the same as 
(\ref{eq:Fconstraint}). Thus the remainder of this
proof mimics that of the previous section. In order for $f$ to
satisfy (\ref{eq:Fconstraint3}) it must be either Gaussian with
respect to the spatial coordinates $\vec{x}_i$, or it must be
polynomial in the distances $x_{1i}$.
But here the polynomial form is ruled out by the simple fact
that dS-invariance of the correlator demands that $f$ behave
under a dilation as
\eq{
  f(\lambda \vec{x}_1,\dots,\lambda \vec{x}_n) 
  = \lambda^{-n\Delta} f(\vec{x}_1,\dots,\vec{x}_n) .
}
Thus $f$ is Gaussian. $\blacksquare$\\

We can quickly extend this result to the case where operators are
inserted in the neighborhood of both asymptotic boundaries of
global dS. It is convenient to switch time coordinate 
to $q = - \ell^2/\eta$, yielding the line element
\eq{
  ds^2 = - \frac{\ell^2}{q^2} dq^2 + \frac{q^2}{\ell^2} d\vec{x}^2 ,
  \quad q \in \Reals.
}
This coordinate chart covers all of $dS_D$. 
In this chart the conformal boundary is composed of two 
copies of $\Reals^d$ located at $|q| \to \infty$, plus two points
located at $q =0$, $|\vec{x}| \to \infty$.
\begin{theorem}\label{thm:Alinear2}
  Let $\phi(x)$ be a scalar operator on $dS_D$, $D \ge 3$,
  with characteristic scaling
  $\phi(x) = O(q^{-\Delta})$, $\Delta > 0$, as $|q| \to \infty$, 
  and let $\Omega$ be a 
  dS-invariant state annihilated by the HS charge $Q_p^{(s)}$.
  Let the action of $Q_p^{(s)}$ on $\phi(x)$ be asymptotically
  linear as $|q| \to \infty$ and contain the term $(p^\mu \d_\mu)^s$.
  Consider the correlation function 
  $\C{\phi(q_1,\vec{x}_1)\dots\phi(q_n,\vec{x}_n)}_\Omega$,
  with each $q_1,\dots,q_n$ equal to $\pm q$, 
  evaluated a non-null separations as $|q| \to \infty$.
  The leading $O(q^{-n \Delta})$ behavior of this
  correlation function is Gaussian.
\end{theorem}
The proof is essentially the same as for the previous theorem.
As for the case of a single boundary, one considers points such
that no $x_{ij} = 0$. This assures that the points $x_i$ and $x_j$
are not coincident (when $q_i = q_j$) or null-separated (when $q_i = - q_j$)
as $|q| \to \infty$.

\section{Discussion} \label{sec:disc} 

In this work we have examined the constraints imposed by the
presence of higher spin symmetries in dS QFTs. 
Our main result was to show that 
if a HS charge acts linearly on a scalar operator near the asymptotic
conformal boundaries, then the vacuum expectation values of that operator
are asymptotically Gaussian. Thus the cosmological spectra associated with
the scalar field are trivial, as is ``scattering'' in global dS.
The condition that a HS charge have asymptotically linear action
is analogous to the stipulation of the Coleman-Mandula theorem that 
symmetries of the S-matrix map $n$-particle states to $n$-particle 
states. Thus, we regard our result as a de Sitter analogue of the 
Coleman-Mandula theorem, specialized to the case of HS symmetries.

Although for simplicity we have focused our discussion on scalar 
operators with real weight $\Delta > 0$,
we expect quite analogous results to hold for spinor and tensor operators,
as well as scalar operators with complex weights. 
It would be interesting to consider vacuum states which do not satisfy
the microlocal spectrum condition (i.e., de Sitter 
``$\alpha$-vacua'' \cite{Mottola:1984ar,Allen:1985ux}).
At the level of linear theories, these states are known to have
interesting scattering properties \cite{Lagogiannis:2011st}; they
may also admit novel interpretations in dS/CFT \cite{Bousso:2001mw}.
Our results rely crucially on the dilation and SCT symmetries
of dS. 
It would therefore also be interesting to examine theories with HS 
symmetry on backgrounds for which the dS symmetry is slightly broken by
non-zero slow-roll parameters.

Finally, we note that our main results,
theorems~\ref{thm:Alinear1} and \ref{thm:Alinear2},
require spacetime dimension $D > 2$. In 2D Minkowski space
there exist a rich class of non-conformal QFTs with HS symmetry
and integrable scattering matrices (see e.g., 
\cite{Goldschmidt:1980wq,Abdalla:1986xb,Evans:2005aa,Lamers:2015dfa}).
It would be very interesting to examine these theories on a dS
background. It is possible that the HS symmetries, if they survive
the relocation to dS, could provide integrable structures and allow
exact results analogues to the Minkowski S-matrices.
We leave this for future study.

\acknowledgments 

We thank Robert Brandenberger, Daniel Harlow, Emil Mottola, 
Gim Seng Ng, Abhishek Pathak and Alexander Zihboedov 
for useful conversations.
We thank the International Institute of
Physics, UFRN, for hospitality while this project was underway.
IM thanks the Kavli Institute for Theoretical Physics, and 
RC thanks McGill University, for additional hospitality.
IM is supported in part by fellowships
from the Institute of Particle Physics, a
Trottier postdoctoral fellowship, and funds
from NSERC Discovery grants.
RC is supported by CAPES (Proc.~5149/2014-02).
This research was supported in part by the National Science Foundation 
under Grant No.~NSF PHY11-25915. 

\appendix

\section{Proof of lemma 3.1}\label{app:lemmas} 

To begin we write the general form of the commutator
of $\phi(x)$ with a HS charge,
\eq{ \label{eq:ansatz}
  \Com{Q_K^{(s)}}{ \phi(x)} = \sum_A 
  \tilde{C}_A^{\mu_1\dots\mu_k}(x) \cO_{\mu_1\dots\mu_k}^A(x) ,
}
where the
$\tilde{C}_A^{\mu_1\dots\mu_k}(x)$ are coefficient functions
which are taken to have length dimension $k-s$;
in order to make the dimensionality
of this equation consistent, the operators $\cO^A_{\mu_1\dots\mu_s}(x)$
must have length dimension $1-D/2-k$.
Consider the action of a dS generator on this commutator.
From the Jacobi identity we obtain
\eq{ \label{eq:jacobi}
  \Com{G_\xi}{\Com{Q_K^{(s)}}{\phi(x)}}
  = \Com{\Com{G_\xi}{Q_K^{(s)}}}{\phi(x)}
  + \Com{Q_K^{(s)}}{\Com{G_\xi}{\phi(x)}} .
}
We say that $G_\xi$ preserves $Q^{(s)}_K$ when
\eq{
  \Com{G_\xi}{Q_K^{(s)}} = \epsilon Q_K^{(s)} ,
}
for some constant $\epsilon$. In this case we obtain from
(\ref{eq:jacobi})
\eq{
  \Com{G_\xi}{\Com{Q_K^{(s)}}{\phi(x)}}
  = \left( i \Lie_\xi + \epsilon\right) \Com{Q_K^{(s)}}{\phi(x)} .
}
On the other hand, from the ansatz (\ref{eq:ansatz}) it follows
that
\eq{
  \Com{G_\xi}{\Com{Q_K^{(s)}}{\phi(x)}}
  = \sum_A 
  \tilde{C}_A^{\mu_1\dots\mu_k}(x) i \Lie_\xi \cO_{\mu_1\dots\mu_k}^A(x) .
}
Taking the difference of these equations we obtain a constraint on 
the coefficient functions:
\eq{ \label{eq:LieC}
  \left( i \Lie_\xi + \epsilon\right) \tilde{C}_A^{\mu_1\dots\mu_k}(x) = 0 .
}

We now apply this result to $Q^{(s)}_p$.
There are several dS generators which preserve this charge.
First consider the translation generators $P_a$ for which $\epsilon =0$.
It follows from (\ref{eq:LieC}) that the
$\tilde{C}_A^{\mu_1\dots\mu_k}(x)$ cannot depend on the spatial variables $x^a$.
The generator of dilations $D$ also preserves $Q_P^{(s)}$, with 
$\epsilon = is$. In this case (\ref{eq:LieC}) requires that 
non-zero components of $\tilde{C}_A^{\mu_1\dots\mu_k}(\eta)$ be $O(\eta^{k-s})$.
For $D \ge 4$ there exist rotations which leave $p^\mu$
unchanged; the associated generators $R^\perp$ preserve
$Q_p^{(s)}$ with $\epsilon = 0$.
The existence of these generators
implies that $\tilde{C}_A^{\mu_1\dots\mu_k}$ are composed of tensors 
invariant under the $SO(d-1)$ rotations which preserve $p^\mu$.

We may also consider the discrete parity transformations of $\Reals^d$ on 
the constant $\eta$ surfaces. Those that preserve $p^\mu$ imply that, 
in all dimensions, the $\tilde{C}_A^{\mu_1\dots\mu_k}(\eta)$ 
are equal to $\eta^{k-s}$ times a constant tensor composed of $\delta^{ab}$, 
$p^\mu$, and $t^\mu = \delta^\mu_\eta$.
The discrete transformation which acts as $p^\mu \to - p^\mu$ further
requires the $\tilde{C}_A$ to have an even (odd) number of $p^\mu$'s
when $s$ is even (odd).
Furthermore, any factor of $\delta^{ab}$ in $\tilde{C}_A^{\mu_1\dots\mu_k}(\eta)$
may be re-cast as factor of the inverse metric $g^{\mu\nu}$, and
this may be absorbed into the definition of the relevant operator.

Bringing everything together, we conclude that we may write
the coefficients as 
\eq{
  \tilde{C}_A^{\mu_1\dots\mu_k}(\eta) 
  = \frac{1}{\eta^{s-k}} C_A^{\mu_1\dots\mu_k},
} 
where the $C_A^{\mu_1\dots\mu_k}$ are constant coefficients composed
of factors of $p^\mu$ and $t^\mu$, with the additional requirement
of $s$ modulo 2 factors of $p^\mu$.
Up to this point we have proven the form 
(\ref{eq:Qphi}), except that we have yet to limit the range of $k$.

Next we consider those dS generators which do not preserve $Q_p^{(s)}$.
In this case the action of a generator on a higher-spin charge 
produces a new charge,
\eq{ \label{eq:GchangeQ}
  \Com{G_\xi}{Q_{K_1}^{(s)}} = \kappa Q_{K_2}^{(s)} ,
}
for some constant $\kappa$. By (\ref{eq:GQ})
$K_2^{\mu_1\dots\mu_s} \propto \Lie_\xi K_1^{\mu_1\dots\mu_s}$. 
We may once again use the Jacobi identity to obtain a constraint 
satisfied by the coefficients functions, though now this constraint 
involves the coefficient functions corresponding to two HS charges. 
For the case (\ref{eq:GchangeQ}) we obtain
\eq{ \label{eq:GchangeQ2}
  -i \Lie_\xi C_{A,K_1}^{\mu_1\dots\mu_k}(x) 
  = \kappa C_{A,K_2}^{\mu_1\dots\mu_k}(x) .
}

In order to use (\ref{eq:GchangeQ2}), let us consider
without loss of generality case $p^\mu \d_\mu = \d_1$, and 
let $s_1^\mu$ be the KVF associated with the special conformal 
transformation with parameter $b^\mu \propto p^\mu$.
It follows from the $SO(D,1)$ algebra satisfied by the KVFs that
\eq{
  \left(\Lie_{s_1}\right)^{2s+1} \left( p^{\mu_1}\dots p^{\mu_s} \right) = 0 .
}
Denoting the SCT generator associated with $s_1^\mu$ by $K_1$ it then
follows that
\eq{
  \Big[ \underbrace{K_1, \Big[ K_1,\dots [K_1}_{2s+1}, Q_{p}^{(s)}]\dots\Big]\Big]
   = 0 .
}
Combining this result with (\ref{eq:GchangeQ}) we obtain the following
constraint on the coefficient functions:
\eq{ \label{eq:SCTconstraint}
  \left(\Lie_{s_1} \right)^{2s+1} 
  \left(\eta^{k-s} C_A^{\mu_1\dots\mu_k}\right) = 0 .
}
Given the form of $C_A^{\mu_1\dots\mu_k}$ this equation is satisfied 
only for $k \le s$. To see this we first note
\eq{
  \Lie_{s_1}^3 \eta^{-1} = 0, \quad
  \Lie_{s_1}^3 p^\mu = 0, \quad
  \Lie_{s_1}^3 t^\mu = 0 ,
}
from which it follows that (\ref{eq:SCTconstraint}) is satisfied
for $0 \le k \le s$.
However, $\Lie_{s_1}$ does not annihilate positive powers of $\eta$,
and thus (\ref{eq:SCTconstraint}) is not satisfied for $k > s$.
This proves the lemma. $\blacksquare$


\addcontentsline{toc}{section}{References}
\bibliographystyle{JHEP}
\bibliography{IAMBib}

\providecommand{\href}[2]{#2}\begingroup\raggedright\begin{thebibliography}{10}

\bibitem{Mukhanov:2005sc}
V.~Mukhanov, {\em {Physical Foundations of Cosmology}}.
\newblock Cambridge University Press, Oxford, 2005.

\bibitem{Coleman:1967ad}
S.~R. Coleman and J.~Mandula, {\it {All Possible Symmetries of the S Matrix}},
  {\em Phys.Rev.} {\bf 159} (1967) 1251--1256.

\bibitem{Parke:1980ki}
S.~J. Parke, {\it {Absence of particle production and factorization of the S
  matrix in (1+1)-dimensional models}},  {\em Nucl.Phys.} {\bf B174} (1980)
  166.

\bibitem{Zamolodchikov:1985wn}
A.~Zamolodchikov, {\it {Infinite Additional Symmetries in Two-Dimensional
  Conformal Quantum Field Theory}},  {\em Theor.Math.Phys.} {\bf 65} (1985)
  1205--1213.

\bibitem{Maldacena:2011jn}
J.~Maldacena and A.~Zhiboedov, {\it {Constraining Conformal Field Theories with
  A Higher Spin Symmetry}},  \href{http://xxx.lanl.gov/abs/1112.1016}{{\tt
  1112.1016}}.

\bibitem{Marolf:2012kh}
D.~Marolf, I.~A. Morrison, and M.~Srednicki, {\it {Perturbative S-matrix for
  massive scalar fields in global de Sitter space}},  {\em Class. Quant. Grav.}
  {\bf 30} (2013) 155023, [\href{http://xxx.lanl.gov/abs/1209.6039}{{\tt
  1209.6039}}].

\bibitem{Maldacena:2002vr}
J.~M. Maldacena, {\it {Non-Gaussian features of primordial fluctuations in
  single field inflationary models}},  {\em JHEP} {\bf 05} (2003) 013,
  [\href{http://xxx.lanl.gov/abs/astro-ph/0210603}{{\tt astro-ph/0210603}}].

\bibitem{Creminelli:2004yq}
P.~Creminelli and M.~Zaldarriaga, {\it {Single field consistency relation for
  the 3-point function}},  {\em JCAP} {\bf 0410} (2004) 006,
  [\href{http://xxx.lanl.gov/abs/astro-ph/0407059}{{\tt astro-ph/0407059}}].

\bibitem{Cheung:2007sv}
C.~Cheung, A.~L. Fitzpatrick, J.~Kaplan, and L.~Senatore, {\it {On the
  consistency relation of the 3-point function in single field inflation}},
  {\em JCAP} {\bf 0802} (2008) 021,
  [\href{http://xxx.lanl.gov/abs/0709.0295}{{\tt 0709.0295}}].

\bibitem{Hinterbichler:2012nm}
K.~Hinterbichler, L.~Hui, and J.~Khoury, {\it {Conformal Symmetries of
  Adiabatic Modes in Cosmology}},  {\em JCAP} {\bf 1208} (2012) 017,
  [\href{http://xxx.lanl.gov/abs/1203.6351}{{\tt 1203.6351}}].

\bibitem{McFadden:2014nta}
P.~McFadden, {\it {Soft limits in holographic cosmology}},  {\em JHEP} {\bf 02}
  (2015) 053, [\href{http://xxx.lanl.gov/abs/1412.1874}{{\tt 1412.1874}}].

\bibitem{Vasiliev:1990en}
M.~A. Vasiliev, {\it {Consistent equation for interacting gauge fields of all
  spins in (3+1)-dimensions}},  {\em Phys.Lett.} {\bf B243} (1990) 378--382.

\bibitem{Anninos:2011ui}
D.~Anninos, T.~Hartman, and A.~Strominger, {\it {Higher Spin Realization of the
  dS/CFT Correspondence}},  \href{http://xxx.lanl.gov/abs/1108.5735}{{\tt
  1108.5735}}.

\bibitem{Hollands:2008vx}
S.~Hollands and R.~M. Wald, {\it {Axiomatic quantum field theory in curved
  spacetime}},  {\em Comm. Math. Phys.} {\bf 293} (2010) 85--125,
  [\href{http://xxx.lanl.gov/abs/0803.2003}{{\tt 0803.2003}}].

\bibitem{Hollands:2014eia}
S.~Hollands and R.~M. Wald, {\it {Quantum fields in curved spacetime}},
  \href{http://xxx.lanl.gov/abs/1401.2026}{{\tt 1401.2026}}.

\bibitem{Wald:1995yp}
R.~M. Wald, {\em {Quantum field theory in curved space-time and black hole
  thermodynamics}}.
\newblock Univ. Pr., Chicago, USA, 1994.
\newblock 205 p.

\bibitem{Mottola:1984ar}
E.~Mottola, {\it {Particle Creation in de Sitter Space}},  {\em Phys. Rev.}
  {\bf D31} (1985) 754.

\bibitem{Allen:1985ux}
B.~Allen, {\it {Vacuum States in de Sitter Space}},  {\em Phys. Rev.} {\bf D32}
  (1985) 3136.

\bibitem{Brunetti:1995rf}
R.~Brunetti, K.~Fredenhagen, and M.~Kohler, {\it {The Microlocal spectrum
  condition and Wick polynomials of free fields on curved space-times}},  {\em
  Comm. Math. Phys.} {\bf 180} (1996) 633--652,
  [\href{http://xxx.lanl.gov/abs/gr-qc/9510056}{{\tt gr-qc/9510056}}].

\bibitem{Radzikowski:1996aa}
M.~J. Radzikowski, {\it Micro-local approach to the hadamard condition in
  quantum field theory on curved space-time},  {\em Comm. Math. Phys.} {\bf
  179} (1996) 529--553.

\bibitem{Brunetti:1999jn}
R.~Brunetti and K.~Fredenhagen, {\it {Microlocal analysis and interacting
  quantum field theories: Renormalization on physical backgrounds}},  {\em
  Commun.Math.Phys.} {\bf 208} (2000) 623--661,
  [\href{http://xxx.lanl.gov/abs/math-ph/9903028}{{\tt math-ph/9903028}}].

\bibitem{Marolf:2010nz}
D.~Marolf and I.~A. Morrison, {\it {Infrared stability of de Sitter QFT:
  Results at all orders}},  {\em Phys. Rev.} {\bf D84} (2011), no.~4 044040,
  [\href{http://xxx.lanl.gov/abs/1010.5327}{{\tt 1010.5327}}].

\bibitem{Hollands:2010pr}
S.~Hollands, {\it {Correlators, Feynman diagrams, and quantum no-hair in
  deSitter spacetime}},  \href{http://xxx.lanl.gov/abs/1010.5367}{{\tt
  1010.5367}}.

\bibitem{Korai:2012fi}
Y.~Korai and T.~Tanaka, {\it {QFT in the flat chart of de Sitter space}},
  \href{http://xxx.lanl.gov/abs/1210.6544}{{\tt 1210.6544}}.

\bibitem{Hollands:2011we}
S.~Hollands, {\it {Massless interacting quantum fields in deSitter spacetime}},
   {\em Annales Henri Poincare} {\bf 13} (2011) 1039--1081,
  [\href{http://xxx.lanl.gov/abs/1105.1996}{{\tt 1105.1996}}].

\bibitem{Rajaraman:2010xd}
A.~Rajaraman, {\it {Proper treatment of massless fields in Euclidean de Sitter
  space}},  {\em Phys. Rev.} {\bf D82} (Dec, 2010) 123522,
  [\href{http://xxx.lanl.gov/abs/1008.1271}{{\tt 1008.1271}}].

\bibitem{Miao:2010vs}
S.~P. Miao, N.~C. Tsamis, and R.~P. Woodard, {\it {de Sitter breaking through
  infrared divergences}},  {\em J. Math. Phys.} {\bf 51} (2010), no.~7 072503,
  [\href{http://xxx.lanl.gov/abs/1002.4037}{{\tt 1002.4037}}].

\bibitem{Prokopec:2003bx}
T.~Prokopec and R.~P. Woodard, {\it {Vacuum polarization and photon mass in
  inflation}},  {\em Am. J. Phys.} {\bf 72} (2004) 60--72,
  [\href{http://xxx.lanl.gov/abs/astro-ph/0303358}{{\tt astro-ph/0303358}}].

\bibitem{Kitamoto:2011yx}
H.~Kitamoto and Y.~Kitazawa, {\it {Infra-red effects of Non-linear sigma model
  in de Sitter space}},  \href{http://xxx.lanl.gov/abs/1109.4892}{{\tt
  1109.4892}}.

\bibitem{Burgess:2010dd}
C.~P. Burgess, R.~Holman, L.~Leblond, and S.~Shandera, {\it {Breakdown of
  Semiclassical Methods in de Sitter Space}},  {\em Journal of Cosmology and
  Astroparticle Physics} {\bf 2010} (2010), no.~10 017,
  [\href{http://xxx.lanl.gov/abs/1005.3551}{{\tt 1005.3551}}].

\bibitem{Bros:2010aa}
J.~Bros, H.~Epstein, and U.~Moschella, {\it Scalar tachyons in the de sitter
  universe},  {\em Lett. Math. Phys.} {\bf 93} (2010) 203--211,
  [\href{http://xxx.lanl.gov/abs/1003.1396}{{\tt 1003.1396}}].

\bibitem{Marolf:2011aa}
D.~Marolf and I.~A. Morrison, {\it {The IR stability of de Sitter QFT: Physical
  initial conditions}},  {\em General Relativity and Gravitation} (2011) 1--34,
  [\href{http://xxx.lanl.gov/abs/1104.4343}{{\tt 1104.4343}}].

\bibitem{Fitzpatrick:2011ia}
A.~L. Fitzpatrick, J.~Kaplan, J.~Penedones, S.~Raju, and B.~C. van Rees, {\it
  {A Natural Language for AdS/CFT Correlators}},
  \href{http://xxx.lanl.gov/abs/1107.1499}{{\tt 1107.1499}}.

\bibitem{SimmonsDuffin:2012uy}
D.~Simmons-Duffin, {\it {Projectors, Shadows, and Conformal Blocks}},  {\em
  JHEP} {\bf 1404} (2014) 146, [\href{http://xxx.lanl.gov/abs/1204.3894}{{\tt
  1204.3894}}].

\bibitem{Smirnov:2004ym}
V.~A. Smirnov, {\it {Evaluating Feynman integrals}},  {\em Springer Tracts Mod.
  Phys.} {\bf 211} (2005) 1--244.

\bibitem{Thompson:1986aa}
G.~Thompson, {\it Killing tensors in spaces of constant curvature},  {\em
  Journal of Mathematical Physics} {\bf 27} (1986), no.~11 2693--2699.

\bibitem{Bakas:1990ry}
I.~Bakas and E.~Kiritsis, {\it {Bosonic Realization of a Universal $W$ Algebra
  and $Z$(infinity) Parafermions}},  {\em Nucl.Phys.} {\bf B343} (1990)
  185--204.

\bibitem{Mikhailov:2002bp}
A.~Mikhailov, {\it {Notes on higher spin symmetries}},
  \href{http://xxx.lanl.gov/abs/hep-th/0201019}{{\tt hep-th/0201019}}.

\bibitem{Lagogiannis:2011st}
P.~Lagogiannis, A.~Maloney, and Y.~Wang, {\it {Odd-dimensional de Sitter Space
  is Transparent}},  \href{http://xxx.lanl.gov/abs/1106.2846}{{\tt 1106.2846}}.

\bibitem{Bousso:2001mw}
R.~Bousso, A.~Maloney, and A.~Strominger, {\it {Conformal vacua and entropy in
  de Sitter space}},  {\em Phys. Rev.} {\bf D65} (2002) 104039,
  [\href{http://xxx.lanl.gov/abs/hep-th/0112218}{{\tt hep-th/0112218}}].

\bibitem{Goldschmidt:1980wq}
Y.~Goldschmidt and E.~Witten, {\it {Conservation laws in some two-dimensional
  models}},  {\em Phys.Lett.} {\bf B91} (1980) 392.

\bibitem{Abdalla:1986xb}
E.~Abdalla, M.~Abdalla, and M.~Forger, {\it {Exact S matrices for anomaly free
  nonlinear sigma models on symmetric spaces}},  {\em Nucl.Phys.} {\bf B297}
  (1988) 374.

\bibitem{Evans:2005aa}
J.~M. Evans, D.~Kagan, C.~A. Young, and N.~J. MacKay, {\it Quantum,
  higher-spin, local charges in symmetric space sigma models},  {\em Journal of
  High Energy Physics} {\bf 2005} (2005), no.~01 020,
  [\href{http://xxx.lanl.gov/abs/0408244}{{\tt 0408244}}].

\bibitem{Lamers:2015dfa}
J.~Lamers, {\it {A pedagogical introduction to quantum integrability, with a
  view towards theoretical high-energy physics}},  {\em PoS} {\bf Modave2014}
  (2015) 001, [\href{http://xxx.lanl.gov/abs/1501.06805}{{\tt 1501.06805}}].

\end{thebibliography}\endgroup

\end{document}